\documentstyle[10pt]{article}

\def\ltap{\raisebox{-.4ex}{\rlap{$\sim$}} \raisebox{.4ex}{$<$}}

\newcommand{\Uthl}{{\rm U(3)}_l}
\newcommand{\Utwl}{{\rm U(2)}_l}
\newcommand{\Uonl}{{\rm U(1)}_l}
\newcommand{\Uthq}{{\rm U(3)}_q}
\newcommand{\Utwq}{{\rm U(2)}_q}
\newcommand{\Uonq}{{\rm U(1)}_q}

\begin{document}
\vspace*{-1in}
\newcommand{\Rslash}{{\not \! \!{R}}}
\renewcommand{\thefootnote}{\fnsymbol{footnote}}
\begin{flushright}
IFUP--TH 31/97\\
\texttt{hep-ph/9707297} \\
To appear in Phys. Rev. {\bf D} \\ 
\end{flushright}
\vskip 5pt
\begin{center}
{\Large{\bf Non-abelian flavour symmetry and $R$-parity }}
\vskip 25pt 
{\bf Gautam Bhattacharyya\footnote{Electronic address:
gautam@mail.cern.ch.~ Permanent address from 1 January 1998: Saha
Institute of Nuclear Physics, \\ 1/AF Bidhan Nagar, Calcutta 700064,
India}}
\vskip 10pt
{\it Dipartimento di Fisica, Universit{\`a} di Pisa 
and INFN, Sezione di Pisa, I-56126 Pisa, Italy }

\vskip 20pt
 
{\bf Abstract}
\end{center}

\begin{quotation}
{\small 
If $R$-parity violation turns out to be a true aspect of Nature, a
speculation about its possible origin could add a new dimension to the
supersymmetric flavour problem. It has been shown in the past by
Barbieri, Hall and their collaborators that the small breaking
parameters of an approximate non-abelian flavour symmetry could govern
the light quark and lepton masses and at the same time could account
for the near degeneracies of squarks and sleptons. A possible
connection of the above feature to the natural suppressions of
$R$-parity-violating couplings has been investigated here. With some
modifications of the approximate flavour symmetry, a supersymmetric
theory without $R$-parity has been motivated that has testable
experimental signatures. \\
PACS number(s): 11.30.Fs, 11.30.Hv, 12.60.Jv}
\end{quotation}

\vskip 20pt  

\setcounter{footnote}{0}
\renewcommand{\thefootnote}{\arabic{footnote}}

Is it possible to reconcile the conventional notion of flavour physics
in supersymmetry concerning masses and mixings and the scenario of
$R$-parity violation? In this paper, we seek for a phenomenologically
viable solution to this question within the framework of a non-abelian
flavour symmetry.  $R$-parity is a discrete symmetry, defined as
$(-1)^{3B+L+2S}$, where $B$ and $L$ are the baryon- and lepton-numbers
and $S$ is the intrinsic spin of a particle \cite{def}. It is $+1$ for
all Standard Model particles and $-1$ for their superpartners. Recall
that neither $L$- nor $B$-conservation is ensured by gauge
invariance. But their uncontrolled violation leads to rapid proton
decay and speeds up many other physical processes at unwanted rates:
these prompted to impose $R$-parity in {\it canonical} supersymmetric
theories. However, violating $R$-parity \cite{wein, hallsuz} in a {\it
controlled} way has rich phenomenological consequences that in recent
times have received considerable attention. An attempt to link
$R$-parity violation to the origin of masses and mixings was made in
the past by invoking a horizontal U(1)-symmetry where charges dictated
by fermion masses and mixings are shown to produce sufficient
suppression in $R$-parity-violating ($\Rslash$) couplings
\cite{BGNN}. Here we are concerned with a non-abelian flavour
symmetry, conjectured first \cite{U2} to realise the conventional
supersymmetric theory of flavour, generalized now to admit $\Rslash$
interactions as well. In addition to maintaining the existing
consistencies and predictions \cite{BHRR,BHR}, our generalization
predicts $\Rslash$ couplings that are within the level of
phenomenological tolerence and lead to detectable signatures. In the
present analysis we consider only the $L$-violating interactions and
leave aside the $B$-violating ones.

In a nutshell, flavour-problem in a supersymmetric theory addresses
the question as how to relate the flavour structure of the fermions
and scalars to each other by the same symmetry principle. An
approximate U(2)-symmetry, which after all descends from a strong
breaking of U(3), through the following step-wise breaking  
\begin{equation}
{\rm U(2)} \stackrel{\epsilon} \rightarrow {\rm U(1)}
\stackrel{\epsilon'} \rightarrow 0,
\end{equation} 
has been shown, in the context of $R$-parity-conserving supersymmetry,
to reproduce the observed patterns of masses and mixings, where
$\epsilon$ and $\epsilon'$ are small dimensionless
breaking-parameters. The three generations of matter fields transform
as $2\oplus 1$, {\it i.e.} $\psi = \psi_a + \psi_3$ ($a=1,2$) and the
`flavon' fields, whose vacuum expectation values (VEVs), after
spontaneous breaking of flavour symmetry, order the mass hierarchies,
have the representations $\phi^a$, $S^{ab}$ (symmetric tensor) and
$A^{ab}$ (antisymmetric tensor). The upper indices in flavons indicate
U(1)-charge opposite to that of $\psi_a$. The first step of breaking
U(2) $\rightarrow$ U(1) is realised through $\langle\phi^2\rangle
\approx \langle S^{22}\rangle \approx {\cal{O}}(\epsilon)M$ (the other
components vanish) and the second step U(1) $\rightarrow 0$ is
achieved by $\langle A^{12}\rangle = -\langle A^{21}\rangle \approx
{\cal{O}}(\epsilon')M$, where $M$ is the cut-off of an effective
theory. The same two small parameters, $\epsilon$ and $\epsilon'$, are
responsible for the near degeneracies of the squarks and slepton
masses, leading to a ``super-GIM'' mechanism. With $\epsilon \simeq
0.02$ and $\epsilon' \simeq 0.004$, all observed masses and mixing
patterns are {\it qualitatively} well understood.

If we now assume that the {\it same} flavour-symmetry is responsible
also for an exact $R$-parity, the strengths of the $\Rslash$
interactions are governed by $\epsilon$ and $\epsilon'$. Do the
magnitudes of $\epsilon$ and $\epsilon'$, dictated by the fermion
masses and mixings, inflict the desirable suppressions to the
$\Rslash$ interactions so as to make the scenario phenomenologically
viable? Before attempting to answer this question, we set up our
notations that we follow hereafter. Recalling that $H_d$ (the Higgs
doublet superfield responsible for the masses of isospin $-1/2$
fermions) and $L$ (the lepton doublet superfield) have identical gauge
quantum numbers, the $\mu H_d H_u$-term in the superpotential can now
be generalized to include 3 more similar terms; in compact notation,
\begin{equation} 
\mu_\alpha L_\alpha H_u ~~(\alpha = 0, i),
\label{mu}
\end{equation} 
where $L_0 \equiv H_d$, $\mu_0 \equiv \mu$ and $L_i$ ($i=1,2,3$)
correspond to the three lepton flavours. One also has the following
trilinear $L$-violating interactions in the superpotential:
\begin{equation} 
\frac{1}{2}\lambda_{ijk} L_i L_j E^c_k + \lambda'_{ijk} L_i Q_j D^c_k,
\label{lam} 
\end{equation} 
where $L_i$ and $Q_j$ are lepton and quark doublet superfields and
$E^c_k$ and $D^c_k$ are charged lepton and down quark singlet
superfields; $i, j, k$ run from 1 to 3. {\it A priori}, without any
suppression ({\it e.g.}~from a horizontal symmetry), the natural
expectation is $\mu_i \sim {\cal{O}} (m_Z)$; $\lambda, \lambda' \sim
{\cal{O}} (1)$ and during electroweak breaking $\langle \tilde{\nu}_i
\rangle \sim {\cal{O}} (m_Z)$. But these overwhelmingly violate the
laboratory upper limits of the neutrino (Majorana) masses \cite{pdg}
(all at 95\% C.L.)
\begin{equation} 
m_{\nu_e} \leq 15~ {\rm eV},~~ m_{\nu_\mu} \leq 170~ {\rm
KeV} ~~{\rm and}~~ m_{\nu_\tau} \leq 24 ~{\rm MeV},
\end{equation}
and overshoot the stringent upper limits (indirect) on various
combinations of $\lambda$- and $\lambda'$-couplings by many orders of
magnitude. The most relevant and stringent constraints are shown in
Table 1 (For an extended list of product couplings, see ref.~\cite{CR}
for example).
\begin{table}[htbp]
\begin{center}
\bigskip
\begin{tabular}{|c|c|c|c|}
\hline
$\mu\rightarrow 3e$ & $\lambda_{1j1}\lambda_{1j2},~ 
\lambda_{231}\lambda_{131}~ \ltap~ 7. 10^{-7}$  & 
$\epsilon_K$ & Im $\lambda'_{i12}\lambda'^*_{i21}~
\ltap~ 8. 10^{-12}$ \\
$\Delta m_K$ & $\lambda'_{i12}\lambda'_{i21}~
\ltap~ 1. 10^{-9}$ & 
$\Delta m_B$ & $\lambda'_{i13}\lambda'_{i31}~
\ltap~ 8. 10^{-8}$ \\
$\mu$Ti $\rightarrow e$Ti & $\lambda'_{1k1}\lambda'_{2k1},~
\lambda'_{11j}\lambda'_{21j}~\ltap~ 5. 10^{-8}$  & 
$K_L \rightarrow \mu e$ & $\lambda'_{1k1}\lambda'_{2k2}~
\ltap~ 8. 10^{-7}$ \\
\hline
\end{tabular}
\caption[] {{\it Upper limits on various product-couplings that scale
as $(\tilde{m}/100~{GeV})^2$, where $\tilde{m}$ is the mass of the
relevant scalar that is exchanged.}}
\end{center}
\end{table}
A way out to have naturally suppressed neutrino masses was suggested
in ref.~\cite{BGNN} through a mechanism that approximately aligns
$\mu_\alpha$ with $v_\alpha$ (the VEVs of the neutral scalars in
$L_\alpha$). A perfect alignment can be achieved if (i) the
supersymmetry-breaking $B_\alpha \propto \mu_\alpha$ and (ii)
$\mu_\alpha$ is an eigenvector of $\tilde{m}^2_{\alpha\beta}$, the
soft scalar mass matrix that arises after supersymmetry breaking; even
though misalignment creeps in through radiative corrections
\cite{Nardi}. Breaking an abelian horizontal U(1) symmetry, with
charges appropriately chosen, was shown \cite{BGNN} to yield
$m_{\nu_\tau} \leq 10$ eV (a hot dark matter candidate) and generate
the $\lambda$- and $\lambda'$-couplings with required suppressions so
as not to violate any experimental constraint.

How does an approximate U(2) symmetry fare to achieve the desired
goal?  Since with a non-abelian horizontal symmetry the theory is much
more constrained than with U(1), the task is much more challenging
and, as we will see below, it faces unavoidable experimental
obstructions, yet gives hints as how to generalize and search for a
plausible solution. The $\Rslash$ bilinear and trilinear terms in the
superpotential can be obtained by appropriately contracting the
superfields appearing in eqs.~(\ref{mu}) and (\ref{lam}) with the
flavons.  Given the flavon representations and the hierarchy of their
VEVs during the step-wise breaking of U(2) down to nothing as
mentioned earlier, the order of magnitude of the $\Rslash$ couplings
are given by (to their leading order)\footnote{All $\Rslash$ couplings
involve flavour indices in the weak basis. For our order of magnitude
estimates, a distinction between the weak basis and the mass basis is
not important.},
\\ $\mu_i$-terms:
\begin{equation} 
\mu_1 \sim 0, ~\mu_2 \sim \epsilon \mu, ~\mu_3 \sim \mu;
\label{muu2} 
\end{equation}
$\lambda_{ijk}$-couplings:
\begin{equation}
\label{lamu2}
(121), (131), (133) \sim 0;~(123), (132), (231) \sim \epsilon';~(232),
(233) \sim \epsilon;~ (122) \sim \epsilon'\epsilon;
\end{equation} 
$\lambda'_{ijk}$-couplings:
\begin{eqnarray} 
 & & (111)', (121)', (131)', (112)', (113)', (133)', (211)', (311)', 
(331)', (313)' \sim 0; \nonumber \\ 
 & & (123)', (132)', (231)', (213)', (321)', (312)' \sim \epsilon';~
(122)', (221)', (212)' \sim \epsilon'\epsilon; \\
 & & (223)', (232)', (233)', (322)', (323)', (332)' \sim \epsilon;~
(222)' \sim \epsilon^2;~ (333)' \sim 1. \nonumber 
\end{eqnarray}   
There are two major phenomenological obstacles in the above
formulation. First, $\langle \tilde{\nu}_\tau\rangle$ and $\mu_3 \sim
m_Z$, while neutrino-neutralino mixings constrain them to be $\ltap
\sqrt{m_{\nu_\tau} m_Z} \ltap {\rm 1 ~GeV}$ (assuming $\mu \sim m_Z$)
and second, $\lambda'_{321}\lambda'_{312} \sim \epsilon'^2 \sim
10^{-5}$ and $\lambda'_{231}\lambda'_{213} \sim \epsilon'^2 \sim
10^{-5}$ exceeding the constraints from $\Delta m_K$ and $\Delta m_B$
(see Table 1) by a few orders of magnitude\footnote{The contribution
to $\epsilon_K$ vanishes as $\lambda'_{i12} = - \lambda'_{i21}$
following from the antisymmetric nature of $A$-flavons.}.

The above difficulties are unrepairable and strongly suggest towards
the consideration of U(3), the ultimate flavour symmetry. However,
U(3) has to be `strongly' broken to account for the heavy top
quark. On the other hand, the failure with U(2) guides us to the
necessity of having an additional suppression factor for the third
generation lepton superfield during U(3) $\rightarrow$ U(2) solving
the `$\mu_3$-problem', that is as well expected to inflict
suppressions in U(2)- and U(1)-breaking parameters curing the
product-couplings' overshooting. So in the lepton sector U(3) needs to
be `weakly' broken. Then how about treating leptons and quarks
differently in flavour-space?\footnote{This is indeed against the idea
of unification, but nevertheless a viable option.}

Following the above line of arguments, we consider the flavour symmetry
$\Uthl\otimes \Uthq$, where lepton and quark superfields transform
under different unitary groups. $\Uthq$ is anyhow strongly
broken to $\Utwq$. The complete breaking configuration is
\begin{equation} 
\Uthl\otimes \Uthq \stackrel{*}\rightarrow \Uthl\otimes \Utwq
\stackrel{\epsilon_{3l}}\rightarrow  \Utwl\otimes \Utwq
\stackrel{\epsilon_l, \epsilon} \rightarrow \Uonl\otimes \Uonq
\stackrel{\epsilon'_l, \epsilon'} \rightarrow 0,
\end{equation}
where `$*$' indicates a strong breaking of $\Uthq$. A triplet flavon
$\tilde{\phi_i}$, with VEV assignments $\langle\tilde{\phi_3}\rangle =
\epsilon_{3l}$, $\langle\tilde{\phi_2}\rangle =
\langle\tilde{\phi_1}\rangle = 0$, breaks $\Uthl$ to $\Utwl$. The
subsequent breaking of $\Utwq$ and $\Utwl$ are assisted by the VEVs of
two different sets of flavon fields (one for quarks and the other for
leptons) which are straightforward three dimensional extensions of the
$\phi$-, $S$- and $A$-fields introduced in the context of a general
U(2) having analogous VEV patterns. For those VEVs related to the
lepton sector we assign a suffix $l$.

Before proceeding further, we must first ensure that the observed
fermion masses and mixings are successfully reproduced. A crucial
assumption at this point is called for that, instead of one pair,
there are two pairs of Higgs doublet superfields.  Considering the two
$H_d$-type Higgs superfields, we assume that one ($H_d^l$) couples
only to leptons and the other ($H_d^q$) only to quarks and there is a
non-trivial mixing between them. The physical state that acquires a
VEV during electroweak breaking is assumed to be the one that
dominantly couples to the leptons and is given by
\begin{equation} 
H_d \simeq H_d^l + \xi H_d^q,
\end{equation} 
while the orthogonal state (assumed too heavy) does not acquire any
VEV. The mass matrices of the charged leptons and the down quarks
assume the following form:
\begin{equation}
{\cal{M}}_l = \left( \begin{array}{ccc}
0 & \epsilon'_l & 0 \nonumber \\
-\epsilon'_l & \epsilon_l & \epsilon_l  \\
0 & \epsilon_l & \epsilon_{3l}
\end{array} \right) v_d, ~~~~~~~~~~~
{\cal{M}}_d = \left( \begin{array}{ccc}
0 & \epsilon' & 0 \nonumber \\
-\epsilon' & \epsilon & \epsilon  \\
0 & \epsilon & 1
\end{array} \right) \xi v_d.
\label{m_l}
\end{equation}
The mixing angle $\xi$ is adjusted as $\xi \approx \epsilon_{3l}
m_b/m_\tau$. Choosing $v_d = v/\sqrt{2} \simeq 174$ GeV (where $v$ is
the standard model VEV), we obtain $\epsilon_{3l} \approx m_\tau/v_d
\simeq 0.01$, $\epsilon_l \approx \epsilon_{3l} m_\mu/m_\tau \simeq
6. 10^{-4}$, $\epsilon'_l \approx \epsilon_l \sqrt{m_e/m_\mu} \simeq
4. 10^{-5}$, $\epsilon \approx m_s/m_b \simeq 0.03$ and $\epsilon'
\approx \epsilon \sqrt{m_d/m_s} \simeq 9. 10^{-3}$.  Note that a
`strong' breaking of $\Uthq$ keeps the values of $\epsilon$ and
$\epsilon'$ the same as in a general U(2)-hypothesis; thus all the
consistencies and observable predictions of the latter related to $B$-
and $K$-physics \cite{BHR} automatically apply to our
scenario\footnote{Indeed, $\mu \rightarrow e \gamma$ is suppressed in
our case by several orders of magnitude compared to its observation
level prediction in U(2)-scenario \cite{BHR}.}. On the contrary, a
`weak' breaking of $\Uthl$ inflicts a suppression of 2 orders of
magnitude in the (33)-element of the charged lepton Yukawa matrix that
is fed to $\mu_3$ and the U(2)- and U(1)-breaking parameters in the
lepton sector; we will see later that quantitatively these fit to our
requirement.  The r\^{o}le of Higgs-mixing is obvious now: despite the
`strong' breaking of $\Uthq$ {\it vis-a-vis} the `weak' breaking of
$\Uthl$, it pulls $m_b$ relative to $m_t$ sufficiently low as to place
it close to $m_\tau$.

Now we are all set to check the consistencies as regards the $\Rslash$
couplings. First, we present the order of magnitude estimates of
$\mu_i$, $\lambda_{ijk}$ and $\lambda'_{ijk}$ (to their leading order)
in the present scenario: \\
$\mu_i$-terms:
\begin{equation} 
\mu_1 \sim 0, ~\mu_2 \sim \epsilon_l \mu, ~\mu_3 \sim \epsilon_{3l} \mu; 
\label{muu3}
\end{equation} 
$\lambda_{ijk}$-terms: 
\begin{equation} 
(121), (131), (133) \sim 0;~ (123), (132), (231) \sim \epsilon'_l
\epsilon_{3l}; ~ (232), (233) \sim \epsilon_l \epsilon_{3l};~ (122)
\sim \epsilon'_l \epsilon_l;
\label{lamu3} 
\end{equation} 
$\lambda'_{ijk}$-terms:
\begin{eqnarray}
\label{lampu3} 
 & & (1jk)', (211)', (231)', (213)', (311)', (331)', (313)' \sim 0;
 (221)', (212)' \sim \epsilon_l \epsilon';~ 
 (233)' \sim \epsilon_l; \nonumber \\
 & &  (222)', (223)', (232)' \sim \epsilon_l \epsilon;~ (321)', (312)' 
 \sim \epsilon_{3l} \epsilon'; (322)', (323)', (332)' \sim \epsilon_{3l}
 \epsilon;~  (333)' \sim \epsilon_{3l}. \nonumber \\ 
\end{eqnarray} 
By putting values of the breaking parameters and comparing the
predictions for the various product-couplings with their experimental
upper limits, we observe that the compatibility has improved
considerably compared to the U(2)-scenario. The prediction
$\lambda'_{321}\lambda'_{312} \sim 7. 10^{-9}$ is in a marginally
tight position with respect to the limit from $\Delta m_K$. But the
entries in the Yukawa matrices are always subject to ${\cal{O}} (1)$
uncertainties that one can exploit to stretch the breaking parameters
for accommodating the above constraint. The $\epsilon_K$-constraint is
trivially satisfied as in the case of a general U(2). The other
constraints (including those which are not listed in Table 1) are
comfortably satisfied\footnote{As a matter of principle, one should
check the consistencies with experimental results by expressing all
$\lambda$- and $\lambda'$-couplings with indices in their physical
basis. But we have checked, as in the U(2)-case mentioned earlier,
that this does not change the conclusions drawn above.}. 

Now we turn our attention to the issue of neutrino mass and its
decay. Neutrino mass arises due to neutrino-neutralino mixings
(photino is irrelevant in the context of neutrino mass) and in the
basis $\left(\tilde{L}^0_{\alpha}, \tilde{H}^0_u, \tilde{Z}\right)$
has the following form ($g_W = g/{2\cos\theta_W}$ and a tilde on a
superfield denotes its fermionic component):
\begin{equation}
{\cal{M}}_n = \left( \begin{array}{ccc}
0_{4\times 4} & \mu_\alpha & g_W v_\alpha \nonumber \\
\mu_\alpha & 0  & -g_W v_u  \\
g_W v_\alpha & -g_W v_u & m_{\tilde{Z}}
\end{array} \right),
\label{m_n}
\end{equation}
where $v_u = \langle H^0_u\rangle$. The zeros in the first ($4\times
4$)-block can be lifted by non-renormalizable terms in the
superpotential of the form $LLH_uH_u/M$, which of course can be
arranged to have a negligible correction assuming $M \gg m_Z$. The
above ($6\times 6$)-matrix has two zero eigenvalues that can be
identified with the physical $\nu_e$ and $\nu_\mu$ masses, while the
physical $\nu_\tau$ is massive and its mass is determined by the
extent to which $v_3$ is misaligned with $\mu_3$ (neglecting, for the
sake of simplicity, the misalignment between $v_2$ and $\mu_2$ which
turns out to be much smaller: recall that with perfect alignment of
all $v_\alpha$ with their corresponding $\mu_\alpha$, all the
neutrinos are massless\footnote{That the three light neutral fermions
(two massless and one massive at tree level) {\it do} correspond to
the three physical neutrinos, is ensured by a simultaneous study of
the charged fermion mass matrix. For a discussion of how to appreciate
this aspect through basis transformations of neutral and charged
fermions, see refs.~\cite{BGNN,Nardi}. In our case, because of the
hierarchical nature of the VEVs of family symmetry breaking, the
neutrino that becomes massive turns out to be {\it dominantly}
$\nu_\tau$. Indeed, higher order effects finally turn the massless
states into massive ones: we ignore those effects here.}). Assuming
for an illustration (good enough for an order of magnitude estimate)
that $B$ is universal and the origin of a possible misalignment is
only an off-diagonal entry $\Delta m^2 = \tilde{m}_{H_d L_3}^2$ in the
scalar lepton mass matrix, an explicit scalar potential minimization
yields
\begin{equation} 
v_3 = \kappa \mu_3 + \kappa' v_d,
\label{potmin}
\end{equation} 
where $\kappa = B v_u/\tilde{m}^2$ and $\kappa' = \Delta
m^2/\tilde{m}^2$ ($\tilde{m}$ is a common diagonal soft scalar
mass). It also follows from the scalar potential minimization that to
a very good approximation $v_d \simeq \kappa \mu$. Therefore, a
non-zero $\kappa'$ is responsible for the deviation from $v_\alpha
\propto \mu_\alpha$ alignment giving rise to a neutrino mass. Now,
$\nu_\tau$-mass is obtained by taking the ratio of the determinant of
the ($4\times 4$) mass matrix [in the ($\nu_\tau$, $\tilde{H}^0_d$,
$\tilde{H}^0_u$, $\tilde{Z}$) basis] to the determinant of the
($3\times 3$) mass matrix [in the ($\tilde{H}^0_d$, $\tilde{H}^0_u$,
$\tilde{Z}$) basis]. The leading behaviour turns out to be
\begin{equation} 
m_{\nu_\tau} \sim \frac{g^2}{4\cos^2\theta_W} \frac{\epsilon_{3l}^2
v_d^2}{m_{\tilde{Z}}},
\label{mnuu3}
\end{equation}
where we have used $\Delta m^2 \approx \epsilon_{3l} \tilde{m}^2$
following from $\Uthl$-breaking. Thus for $m_{\tilde{Z}} \sim v_d$,
$m_{\nu_\tau} \sim {\cal{O}} (1~{\rm MeV})$ lying in the range of
detectability, for example, at a tau-charm factory \cite{taucharm}.

However, this massive $\nu_\tau$ is not stable and before we discuss
its decay properties, a few remarks on the cosmological constraints
that apply on it are in order \cite{reviews}. The age and the present
energy density of the universe restricts the lifetime of a 1 MeV
$\nu_\tau$ to be less than $\sim 10^8 s$. A stronger constraint
(lifetime less than $\sim 10^3 s$) follows from the requirement that
$\nu_\tau$ should decay before the recombination time ($t_{\rm rec}
\ltap 10^{-5} t_U$, where $t_U$ is the age of the universe being
$10^{10} y$), {\it i.e.}~when matter could start forming.  The
nucleosynthesis upper bound on the lifetime of a 1 MeV neutrino is
$\sim 10^2 s$, unless it has additional annihilation channels besides
those in the Standard Model. When the dominant decays are in visible
channels ({\it e.g.}~radiative decays), practically all otherwise
allowed neutrino masses are excluded\footnote{See {\it e.g.}  Fig. 2
of Gelmini and Roulet in ref.~\cite{reviews}.}.

Within our framework, $\nu_\tau$ has three types of decay modes: \\
(i) Invisible decay $\nu_\tau \rightarrow \nu_\mu f$, where $f$ is a
familon \cite{wilczek,gel} (a massless Nambu-Goldstone boson arising
from the breaking of the family symmetry $\Uthl$).  The effective
operator $LLH_uH_u/M$ induces this decay (recall that a familon does
not carry any overall lepton number) and the loop-driven decay graph
involves two $\Rslash$ Yukawa couplings ({\it e.g.} $\lambda'_{333}$
and $\lambda'_{233}$) generating $\Delta L = 2$; \\ (ii) Invisible
decay to three light neutrinos, $\nu_\tau \rightarrow 3\nu$
($Z$-mediated), following from the frustration of GIM-mechanism due to
neutrino -- zino mixing\footnote{Charged lepton -- chargino mixing
will trigger flavour-changing Z decays into light leptons, $Z
\rightarrow l_i \bar{l}_j$, the rates of which, we have checked, are
much below their experimental upper limits \cite{pdg}.}; \\ (iii)
Visible radiative decay $\nu_\tau \rightarrow \nu_\mu + \gamma$,
induced by $\lambda'_{333}$ and $\lambda'_{233}$ (for example).\\ For
superparticle masses around 100 GeV, the lifetime in channel (i) is
$\sim 10^{16} s$ with $V \sim 6. 10^9$ GeV (global $\Uthl$ breaking
scale\footnote{This lower limit follows from the non-observation of
the $\mu \rightarrow e f$ decay \cite{ambro}.})  while the lifetimes
in channels (ii) and (iii) are $\sim 10^{12}$--$10^{13} s$. It should
be noted though that the lack of finding a fast enough decay channel
of a massive neutrino is a somewhat generic problem that has been
noticed in the past in different contexts \cite{reviews,BrGNN}. We
observe that we cannot advance any solution to this general problem in
a scenario where approximate non-abelian horizontal symmetries have
been assumed to control {\it both} the $\Rslash$ Yukawa couplings and
the structure of the supersymmetry breaking soft terms.

If we instead assume that family symmetries govern {\it only} the
Yukawa couplings through their hierarchical breaking and do not
control the structure of the soft masses at the supersymmetry breaking
scale ($\Lambda_U$), this indeed results in a loss of generality. But
this is aimed to avoid the difficulties related to the rather long
lifetime of the massive neutrino by bringing its mass below 100 eV
making it cosmologically stable \cite{reviews}. Let us assume the
following: (i) soft terms are universal at $\Lambda_U$, {\it i.e.} 
$\tilde{m}^2_{\alpha\beta} = \tilde{m}^2 \delta_{\alpha\beta}$, (ii)
$B_\alpha = B \mu_\alpha$ and finally (iii) the supersymmetric $\mu$
parameter is non-zero in only one direction, namely, $\mu_\alpha
L_\alpha H_u \equiv \mu H_d H_u$: this is not unjustified as there is
an in-built distinction between $H_d$ and $L_i$, since the former is a
singlet under family group while the latter transforms under
$\Uthl$. Assumption (iii) therefore relies on a property of the theory
that its superpotential could sense that distinction and chooses the
`singlet direction' for the $\mu$-term. Still a question remains: even
if one starts with a universal boundary condition on the scalar masses
at $\Lambda_U$, how much sneutrino-Higgs mixing is generated by
renormalization group (RG) running of the soft parameters down to low
energy? Singling out the dominant effects, an approximate
(nevertheless quite reasonble for an order of magnitude estimate)
expression of the mass of $\nu_\tau$ induced by such misalignment is
obtained as \cite{Nardi,Anjan}
\begin{equation} 
m_{\nu_\tau}^{\rm RG} \sim \frac{g^2}{4\cos^2\theta_W} \frac{
v_d^2}{m_{\tilde{Z}}} \left[\frac{3t_U m_b}{8\pi^2 v}\right]^2
\left(3+\frac{A^2}{\tilde{m}^2}+\frac{A}{B}\right)^2 \lambda'^2_{333},
\label{mnurg}
\end{equation} 
where $t_U = \ln(\Lambda_U/m_Z)$ and $A$ is the universal trilinear
soft parameter at $\Lambda_U$.  By comparing eqs.~(\ref{mnuu3}) and
(\ref{mnurg}) one obtains an idea of the relative sizes of the
RG-induced effect on the neutrino mass and the $\Uthl$-breaking
contribution discussed earlier. Let us consider, for the sake of
simplicity and illustration, $A \ll \tilde{m}, B$. Then, (i) for
$\Lambda_U = 10^{16}~{\rm GeV}$, $m_{\nu_\tau}^{\rm RG}$ is at the
level of a few KeV and (ii) for $\Lambda_U = 10^{5}~{\rm GeV}$,
$m_{\nu_\tau}^{\rm RG}$ is ${\cal{O}} (100~{\rm eV})$. In case (i),
even by exploiting the ${\cal{O}} (1)$ uncertainty in
$\lambda'_{333}$, it is difficult to bring the neutrino mass below 100
eV for natural choices of soft parameters, while in case (ii), which
corresponds to low energy gauge-mediated supersymmetry breaking
\cite{gmsb}, there is more breathing space to accomplish it mainly
because of less RG-running\footnote{In gauge-mediated models, the soft
masses are not universal but flavour symmetric and so our conclusions
remain unaffected.}. At this level it becomes important to evaluate
the one-loop contribution to the neutrino mass induced by (dominantly)
the $\lambda'_{333}$ coupling. The leading term reads \cite{GRT}
\begin{equation} 
m_{\nu_\tau}^{\rm loop} \approx \frac{3m_b m^2_{\rm
LR}}{8\pi^2\tilde{m}^2} \lambda'^2_{333},
\end{equation} 
where assuming the left-right squark mixing $m^2_{\rm LR} = m_b
\tilde{m}$, we obtain, for $\tilde{m} = 100 ~{\rm GeV}$,
$m_{\nu_\tau}^{\rm loop} \sim 1 ~{\rm KeV}$. Again, it is possible to
arrange the squark masses and mixings and/or $\epsilon_{3l}$-scaling such
that $m_{\nu_\tau}^{\rm loop}$ becomes ${\cal{O}} (100~{\rm eV})$. It is
noteworthy that for low energy supersymmetry breaking
$m_{\nu_\tau}^{\rm RG}$ becomes comparable or even less than
$m_{\nu_\tau}^{\rm loop}$, while for $\Lambda_U \sim 10^{16}~{\rm
GeV}$ the dominant contribution comes from misalignment. In any case,
we have exhibited that it is possible to design a scenario
(particularly with gauge-mediated supersymmetry breaking) reconciling
$R$-parity violation with conventional flavour physics that, in
addition to having passed the laboratory tests, is also cosmologically
viable.

In the scenario discussed above, the cosmologically stable neutrinos
are hot dark matter candidates. Axions, that have resulted from
breaking non-abelian, continuous and global family symmetries, could
constitute cosmologically interesting cold dark matter \cite{U2}. The
other candidates for cold dark matter in $R$-parity-conserving
supersymmetry are neutralinos, which are not stable here in
cosmological scales.  Given the predictions of the $\Rslash$-couplings
in eqs.~(\ref{lamu3}) and (\ref{lampu3}), the most striking collider
signatures of this scenario are: (i) [if the lightest neutralino is
the lightest supersymmetric particle (LSP)] like-sign di-muon final
states \cite{GR} from LSP-decays after a rather long flight ($\sim$ 1
m) close to the detector edge and (ii) [in the sneutrino-LSP scenario]
$\tilde{\nu}_\tau$ decaying to 2 jets inside the detector through
$\lambda'_{3ij}$ couplings \cite{BKP}. We note in passing that the
particular couplings ($\lambda'_{1j1}$) relevant to explain the recent
HERA anomaly \cite{hera} are vanishing in our case and so if those
anomalous events turn out to be real in future, they cannot be
explained within our framework. In any case, if $R$-parity-violation
turns out to be a true feature of Nature, we believe that its possible
ancestral link with masses and mixings could constitute a complete
theory of flavour. Our effort is an attempt in that direction.

\vskip 5pt
\noindent
I thank Riccardo Barbieri for suggesting the problem and for helping
and encouraging me at every stage of the work, Andrea Romanino for
a critical reading of the manuscript and Rabi Mohapatra for a
discussion on cosmological constraints on neutrino decays.


\end{document}